\def\kpc{ {\rm kpc} }
\def\pc{ {\rm pc} }

\def\msun{{\rm M}_{\rm \odot}}

\def\sech{{\rm sech}}
\def\zthin{z_{\rm thin}}
\def\zthick{z_{\rm thick}}

\def\spose#1{\hbox to 0pt{#1\hss}}
\def\lta{\mathrel{\spose{\lower 3pt\hbox{$\sim$}}
    \raise 2.0pt\hbox{$<$}}}
\def\gta{\mathrel{\spose{\lower 3pt\hbox{$\sim$}}
    \raise 2.0pt\hbox{$>$}}}
\documentstyle[12pt,aasms4,epsf,rotate]{article}

\begin{document}

\title{The LMC Microlensing Events: Evidence for a Warped and Flaring 
Milky Way Disk?}

\author{N. Wyn Evans}
\affil{Theoretical Physics, Department of Physics, 1 Keble Rd, Oxford,
OX1 3NP, UK}
\author{Geza Gyuk}
\affil{S.I.S.S.A., via Beirut 2-4, 34014 Trieste, Italy}
\author{Michael S. Turner}
\affil{Departments of Astronomy \& Astrophysics and of Physics, Enrico 
Fermi Institute, The University of Chicago, Chicago, IL 60637-1433, USA}
\affil{NASA/Fermilab Astrophysics Center, Fermi National Accelerator 
Laboratory, Batavia, Illinois 60510-0500, USA}
\and
\author{James Binney}
\affil{Theoretical Physics, Department of Physics, 1 Keble Rd, Oxford,
OX1 3NP, UK}

\begin{abstract}{
The simplest interpretation of the microlensing events towards the
Large Magellanic Cloud detected by the MACHO and EROS collaborations
is that about one third of the halo of our own Milky Way galaxy exists
in the form of objects of around $0.5$ solar mass. There are grave
problems with this interpretation.  A normal stellar population of
$0.5$ solar mass stars should be visible. The other obvious candidate
for the lenses is a population of white dwarfs. But, the precursor
population must have polluted the interstellar medium with metals, in
conflict with current population II abundance. Here, we propose a more
conventional, but at the moment more speculative, explanation. Some of
the lenses are stars in the disk of the Milky Way. They lie along the
line of sight to the LMC because of warping and flaring of the
Galactic disk. Depending on its scalelength and ellipticity, the
disk's optical depth may lie anywhere between $0.2 \times 10^{-7}$ and
$0.9 \times 10^{-7}$. Together with contributions from the LMC disk
and bar and perhaps even intervening stellar contaminants, the total
optical depth may match the data within the uncertainties.
Microlensing towards the LMC may be telling us more about the
distorted structure and stellar populations of the outer Milky Way
disk than the composition of the dark halo.}
\end{abstract}


\section{INTRODUCTION}

The MACHO and EROS experiments are monitoring the lightcurves of
millions stars in the nearby Large Magellanic Cloud (LMC) searching
for examples of gravitational microlensing by massive compact objects
that might comprise the halo of our own Galaxy~(\cite{one,three}).
Recently, the MACHO group has announced the provisional discovery of
six more events~(\cite{nine}), taking the total to fourteen events
over the past four years. The tentative estimate of the optical depth
to microlensing is about $2 \times 10^{-7}$ with an uncertainty of
perhaps $1 \times 10^{-7}$. This result must be treated with some
reserve as it is based both on a preliminary event set and on the
efficiency curves calculated from the two-year data-set. The optical
depth may fall still lower once the four-year efficiencies become
available.  If the lenses lie in the halo, then the typical mass of a
lens is around $0.5$ solar masses. The lensing population is
responsible for roughly a third of the overall mass of the halo within
fifty kiloparsecs~(\cite{three}). These innocent-looking statements do
not sit comfortably with other well-established astronomical
facts. First, it is hard to understand what kind of astrophysical
objects comprise the lensing population, as normal stars, red dwarfs
and white dwarfs are not believable~(\cite{four,five,seven,eight}).
Primordial black holes remain possible, although this suggestion faces
a well-known tuning problem~(\cite{carr}).  Second, if this simple
interpretation were correct, it would indicate that the bulk of the
dark baryons are stars, whereas there is ample evidence that as late
as a redshift of one half, most of the baryons were still in gaseous
form (the amount of hot gas in clusters is at least as large as the
maximum abundance of baryons permitted by primordial nucleosynthesis).
Third, the simple interpretation implies that our dark halo has two
different but comparable components -- dark stars and something else,
with particle dark matter as the leading possibility -- whereas
Occam's razor would prefer a single component.

\section{THE WARPED AND FLARING DISK OF THE MILKY WAY}

The aim of this Letter is to suggest a more conventional, albeit
speculative, possibility: Many of the lenses lie in the warped and
flaring disk of the Milky Way and are an ordinary stellar
population. Together with contributions from other foreground stellar
populations, they provide the microlensing optical depth detected by
the observers. In particular, Sahu (1994) pointed out that some of
the lenses may be in the stellar disk and bar of the LMC itself.  The
optical depth to microlensing of the LMC disk and bar is $0.5
\times 10^{-7}$ at most~(\cite{three,eleven}). Gates et al. (1998)
have put forth the possibility that hitherto undetected stellar
populations in the spheroid or thick disk may contribute to the
microlensing optical depth.  Zaritsky \& Lin (1997) have associated a
feature on the colour-magnitude diagram with a possible stellar
population lying some 15 kpc in front of the LMC in a band of material
associated with the Magellanic Stream. They have estimated its optical
depth may be of the order $0.5 - 1.0 \times 10^{-7}$, although others
have contested this interpretation~(\cite{fourteen}).  Our suggestion
is that the Milky Way disk may provide a contribution of up to $0.9
\times 10^{-7}$.  Previous investigators estimated the thin disk to
have an optical depth of $0.15 \times 10^{-7}$ by idealising it as a
flat exponential disk of uniform scale-height~(\cite{three}). This is
likely to be a serious underestimate, as the Milky Way disk shows
very substantial deviations from flatness.  For example, the layer of
neutral hydrogen gas is severely warped beyond the solar
circle~(\cite{seventeen}). The midplane of the gas and the stellar
disk falls to 1.0 - 1.5 kpc below the Galactic equator in the
south~(\cite{nineteen}).  Beyond 15 kpc, it bends back to the
equatorial plane in the south, while continuing to rise to more than 5
kpc out of plane in the north.  The Milky Way disk also flares
strongly beyond the solar circle. The scale-height of the neutral
hydrogen increases by a factor of ten on moving outwards from the
solar circle to Galactocentric radii about 25 kpc~(\cite{twenty}).

Figure 1 shows the idea put forward in this paper in pictorial
form. The warping and flaring of the Milky Way stellar disk can
enhance the density of material along the line of sight towards the
LMC.  To estimate the importance of this effect, let us model the
Milky Way disk as a declining exponential in Galactocentric radius
$R$.  The vertical profile of the thin disk follows roughly a
hyperbolic-secant squared profile whereas the thick disk is
well-modelled by an exponential profile in height above or below the
midplane of the warp $|z-z_w|$. A reasonably realistic density law --
at least locally -- is~(\cite{twentyone})
\begin{equation} 
\rho(R,z) = \rho_0 \exp (-R/ R_d) \Bigl[ (1-f) \exp (-|z - z_w|/
\zthick ) + f \sech^2 ( |z - z_w|/ \zthin ) \Bigr].  
\end{equation} 
Both the warp $z_w$ and the flares of the thin and thick disks
$\zthin$ and $\zthick$ vary with Galactocentric radius.  The
fractional contributions of the thin and thick disk to the midplane
density are denoted by $f$ and $1-f$.  At outset, let us emphasise
that though some of the important parameters defining the Milky Way
disk are securely established, others are much more doubtful.  To give
the reader a feeling for the likely uncertainties, we shall use the
basic density law (1) to generate three sets of models that are
possible representations of the Milky Way disk. At least, the overall
normalisation $\rho_0$ is reasonably well-known. We choose it to
recover the column density of the thin and thick disks at the solar
radius ($R=8\,$kpc) of $42 \msun \pc^{-2}$.  This is slightly higher
than the recent determination from analysis of Hubble Space Telescope
(HST) fields~(\cite{twentyone}), but entirely consistent with
dynamical estimates. For example, Gould (1990) reckons the total mass
of the disk is $54 \pm 6 \msun \pc^{-2}$ of which $13 \msun \pc^{-2}$
is in gas. The remainder is in the stars of the thin and thick disk
and available for microlensing. Over the region of interest, the warp
$z_w$ is approximated by a single sine curve with an amplitude that
rises linearly with distance from the Solar circle and reaches 1.5 kpc
at a Galactocentric radius of 15 kpc.  The scale-height of the thin
disk $\zthin$ at the Sun is 350 pc, whereas the scale-height of the
thick disk is 700 pc~(\cite{twentyone}).  Both the thin and the thick
disk are allowed to flare linearly so that the scale-height increases
by a factor of ten by a Galactocentric radius of 25 kpc.  Our first
set of models (labelled A in Table 1) have a fractional contribution
of $25 \%$ thick disk and $75 \%$ thin disk, as suggested by the most
recent HST data~(\cite{twentyone}). Our second set of models (labelled
`B') are identical except for a differing relative contribution of $10
\%: 90 \%$ for the thick and thin disks. This is the older,
traditional normalisation that predates the recent HST work. The third
set of models (labelled `C') are identical to A, but have a more
modest flare -- the scale-height increases by a factor of five rather
than ten as the distance increases threefold.  Of all the poorly
established Galactic parameters, the disk's radial scale-length $R_d$
is the most significant for the microlensing observables. The
scale-length is different for different stellar populations but
probably lies within the range 2.5 to 5.5
kpc~(\cite{twentythree,twentyfour,twentyfive}).  Rather than mediate
between the different claims of different data-sets, Table 1 presents
results for the broad range of possible scale-lengths suggested by the
observations. The microlensing optical depth is highest for the
strongly flared models with the higher fractional contributions from
the thick disk. If the Milky Way disk is circular, its optical depth
may be as high as $0.64 \times 10^{-7}$. If the Milky Way is slightly
elliptical with the Sun lying on the minor axis -- as has been
suggested on dynamical grounds~(\cite{twentyseven}) -- then the
optical depth may be as high as $0.88 \times 10^{-7}$.

Let us remark that equation (1) describes a model with an exponential
decline in the midplane density. As the scale-height depends on
radius, the column density increases somewhat on moving outward from
the solar circle. Face-on spiral disks often have column densities
that are reasonably well-fit by exponential laws within the galaxy's
optical edge.  This however has the status of a rule-of-thumb, rather
than an immutable physical law. In fact, examination of the infra-red
light profiles (which trace the mass better than the visible) of large
samples of of field spirals shows that deviations of up to 50\% from
best fit exponentials are quite
common~(\cite{twentyeight}). Additionally, severe lopsidedness
($>20\%$) and some degree of ellipticity ($10\%$) are frequent in
late-type spirals~(\cite{twentyeight,twentynine}).  It is also the case
that merging and interacting galaxies can show very large deviations
from simple exponential disks.  While the Milky Way is not presently
undergoing a major merger, the estimated mass of the Sagittarius dwarf
galaxy continues to increase~(\cite{thirty}), leading one to believe
that the damage it has inflicted on the Milky Way disk may be larger
than originally suspected. Numerical simulations of sinking
satellites~(\cite{thirtyone}) show that a very characteristic effect of
accretion is the flaring of fragile galactic disks.  This raises the
possibility that the optical depth of the Milky Way disk might be
still higher than the values reported in Table 1. Simple exponential
disk models may not be an accurate guide to the outer parts.

Alone, the values in Table 1 are lower than the observations.  Taken
together with the contribution of the LMC itself~(\cite{ten}) and the
foreground population~(\cite{thirteen}), the optical depth may match or
even exceed the observations. It is also possible that the experimental
efficiency is imperfectly calibrated -- as is suggested by the serious
discrepancy between observational and theoretical estimates of the optical
depth towards the Galactic Bulge (e.g., Evans 1997, Bissantz et al.
1997). Severe blending (c.f., Goldberg \& Wo\'zniak 1998) may be a cause
of this problem.  If the same problem afflicts the calibration towards the
LMC, then the true optical depth may be lower than observed by a factor of
two.  A warped and flaring disk always makes a substantial contribution to
the optical depth if the fractional normalisation of the thick disk is
$\gta 25 \%$ and if the scaleheight of the stellar warp increases by the
same factor as the scaleheight of the neutral gas. Figure 2 shows the
cumulative probability histogram of the MACHO four-year
data~(\cite{three,nine}) of thirteen events (the binary event has been
removed because it is almost certainly due to LMC self-lensing). Note that
the data shows strong evidence of clumping, with over half the events
possessing timescales between 30 and 45 days. This emerges naturally in
our picture, as the lenses share the velocity field of the rotationally
supported disk and have heliocentric distances of rougly $\sim 20$ kpc.
Thus the mass distribution dominates the observed scatter in the
timescales.  Let us assume that the present-day mass function (PDMF) of
the lenses is a Salpeter-like power-law down to $\sim 0.6\,\msun$ before
flattening off. This is the mass function determined from recent HST
star-count data in high latitude fields~(\cite{twentyone}). The velocity
distribution of the lenses is taken as an anisotropic Gaussian about the
(asymmetric drift corrected) circular speed. Correcting for the efficiency
of the experiment with the two-year efficiency curve~(\cite{three}), our
model predicts that the number of events the MACHO group would have seen
over the four-year period of the experiment is $4-6$. The probability of
detecting an event with timescale less than $t_0$ in our model is plotted
in Figure 2. It is overlaid on the actual timescale histogram of
events. Although the fit is very good, we caution that this is not a
definitive test, as halo models with sharply peaked mass functions can
also give good fits.

\section{DISCUSSION AND CONCLUSIONS}

Our model makes bold claims and is not difficult to test.
Gould (1998) has pointed out that the surface brightness map of the
LMC made by de Vaucouleurs (1957) provides strict constraints on
foreground structures acting as microlenses. The last detectable LMC
isophote is at $25$ mag arcsec${}^{-2}$, so one possibility is for any
intervening material to be fainter than this limit. However, de
Vaucouleurs removed a smooth foreground component with a mean surface
brightness of $21.2$ mag arcsec${}^{-2}$ and a mean gradient of $15
\%$ before constructing the isophotal map. So, any foreground
component that is smooth on the scale of the map (roughly $15^\circ$)
will also have been removed. Figure 3 shows contours of surface
brightness in a Hammer-Aitkof projection of our warped and flaring
disk around the Galactic anti-center. A value of the mass-to-light
ratio of $3$ has been assumed -- quite reasonable for disk
populations.  It is readily seen that the surface brightness contours
have negligible gradient in longitude, while the gradient in latitute
at the LMC's position is below the $15 \%$ removed by de
Vaucouleurs. The surface brightness map therefore cannot rule out
smooth, extended foreground structures such a warped, flaring disk.
One of the best ways of testing our model is using star counts, but it
is important to appreciate that the data must be deep enough to probe
the large-scale structure of the Milky Way disk. It is only recently
that star count data-bases extending far enough from the solar
position have been compiled. One straighforward way to test our model
is to look for evidence of a stellar flare towards the Galactic
anti-center. This is a conclusive test as the microlensing optical
depth in our model depends strongly on the magnitude of the flare.
Integrated starlight measurements may also be useful in elucidating
the global structure of the Milky Way disk but are unfortunately
bedevilled by uncertainties in the distribution of Galactic
extinction. Directly observing the lenses may be possible. However,
the typical distance modulus for lenses is around $17$ making this
very challenging. A more promising possibility is to use the new large
telescopes (such as the VLT) to detect stars in our warped disk.
Although any estimate is necessarily somewhat model-dependent, we
anticipate that probing to 22nd or 23rd magnitude in the LMC direction
and looking for variations in the shape of the starcounts as a
function of magnitude may be a useful diagnostic.  Our model predicts
the optical depth depends quite strongly on Galactic latitude. The LMC
does not span much latitude itself, so the events should be spread
over the whole area of the LMC disk. Towards the higher latitude Small
Magellanic Cloud (SMC), this model predicts a much lower optical depth
of $2.0 \times 10^{-8}$.  A test of the ratios of the optical depths
to the LMC and SMC would be instructive, although the self-lensing of
the SMC may need to be accurately modelled~(\cite{thirtytwo}). Since
the lenses are primarily low-mass disk stars at large distances from
the Sun, neither blending~(\cite{thirtythree}) nor
parallax~(\cite{thirtyfour}) effects are important.

Microlensing began as an innovative approach to identifying the nature
of dark matter in our Galaxy.  Our suggestion that some of the lenses
responsible for the LMC microlensing events may lie in the warped and
flaring disk is speculative, but it provides a coherent explanation of
data that is otherwise frustratingly hard to understand.  If correct,
microlensing may end up teaching us more about the structure of the
Galaxy than about dark matter.  The high microlensing rate toward the
Galactic Center has already provided further confirmation of the
bar-like geometry of the Galactic bulge~(\cite{thirtyfive}).
Similarly, microlensing towards the LMC may be a valuable probe of the
shape and stellar populations of the outer Galactic disk.  Perhaps
more importantly, microlensing will have exhausted the last baryonic
candidate for the halo dark matter, leaving elementary particles left
over from the earliest moments as the likeliest remaining possibility.

\acknowledgments

We thank participants at the Aspen Center for Physics workshop on
microlensing -- especially Tim Axelrod, Eamonn Kerins, David Graff,
Andy Gould, Kris Stanek and Dennis Zaritsky -- for helpful discussion
of this idea.

\eject

\begin{table*}
\begin{center}
\begin{tabular}{crrrrrrrrrrr}
Model&Ellipticity& $2.5\,\kpc$ & $3.5\,\kpc$ & $4.5\,\kpc$ & $5.5\,\kpc$ & \\
\tableline
\null &\null&\null&\null&\null&\null\\
A &1. &$0.300 \times 10^{-7}$ & $0.394 \times 10^{-7}$  
&$0.508 \times 10^{-7}$ & $0.637 \times 10^{-7}$ \\
A &0.85 & $0.395 \times 10^{-7}$ & $0.537 \times 10^{-7}$ & $0.701 \times
10^{-7}$ & $0.877 \times 10^{-7}$ \\
\null &\null&\null&\null&\null&\null\\
B &1. &$0.194 \times 10^{-7}$  &$0.238 \times 10^{-7}$  
&$0.292 \times 10^{-7}$ &$0.354 \times 10^{-7}$ \\
B &0.85 &$0.239 \times 10^{-7}$ & $0.306 \times 10^{-7}$ & $0.385 \times
10^{-7}$& $0.470 \times 10^{-7}$ \\
\null &\null&\null&\null&\null&\null\\
C &1. &$0.206 \times 10^{-7}$ & $0.222 \times 10^{-7}$ 
&$0.241 \times 10^{-7}$ &$0.261 \times 10^{-7}$ \\
C &0.85 &$0.227 \times 10^{-7}$  & $0.249 \times 10^{-7}$  
& $0.275 \times 10^{-7}$ & $0.302 \times 10^{-7}$ \\
\null &\null&\null&\null&\null&\null\\
\end{tabular}
\end{center}


\tablenum{1}
\caption{This table gives the optical depth to microlensing towards
the Large Magellanic Cloud for a sequence of models that span the
possible structure of the Milky Way disk. For each model, the
variation of the optical depth with the scale-length of the disk is
shown along the rows of the table. Model A has the thick and thin 
disks normalised in the ratio $25 \% : 75 \%$, together with a
flare whose scale-height increases by a factor of ten between the
solar circle and $25\,\kpc$. Model B differs from model A only
in the ratio of normalisation of thick to thin disk, which is
taken as $10 \%: 90 \%$. Model C differs from model A in possessing
a more modest flare -- the scale-height is permitted to increase only
by a factor of five over the same radial distance. Results for all
models are given for round ($q =1$) and elliptical ($q=0.85$) Milky
Way disks.}

\end{table*}

\eject

\begin{figure}
\begin{center}
               \epsfxsize 0.43\hsize
               \leavevmode\epsffile{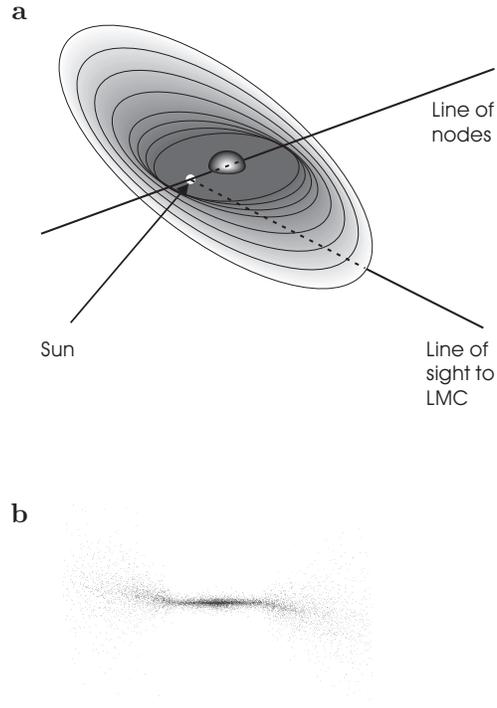}
\caption{This is a schematic figure of the Milky Way stellar disk, 
together with the line of sight from the Sun towards the Large
Magellanic Cloud.  The upper panel (a) shows the behaviour of the
mid-plane of the stellar disk. The Sun lies nearly on the line of
nodes of the warp. The stellar disk is assumed to warp in a linear
fashion to large Galactocentric radii. Note that this is different
from the behaviour of the Milky Way gas disk, which bends back towards
the Galactic equator in the south. The lower panel (b) is a
cross-section along the ridge-line of the warp showing the flare at
large radii. The flare is taken as varing linearly in amplitude just
beyond the solar circle.}
\end{center}
\end{figure}

\begin{figure}
\rotate[r]{
               \epsfxsize 0.7\hsize 
               \epsffile{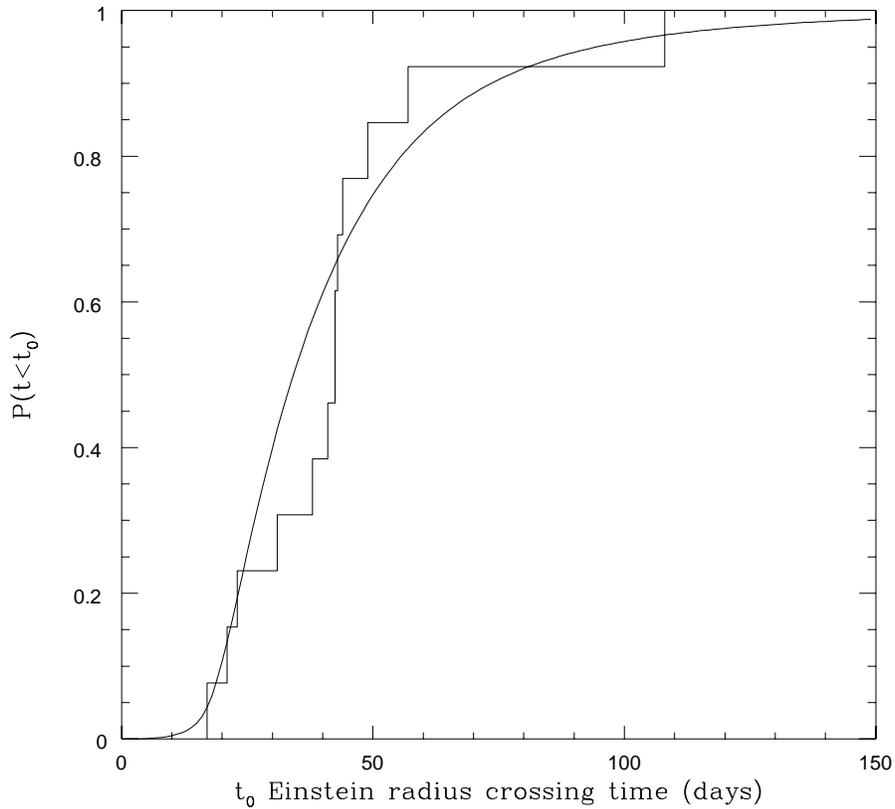}
}
\caption{The probability of detecting an event with timescale less
than $t_0$ is plotted against $t_0$ for our warped, flaring disk
model. This is calculated including the effects of the proper motion
of the LMC, the dispersion and the asymmetric drift of the warped and
flaring stellar population.  However, the effects of the stellar
populations in the Large Magellanic Cloud itself or any intervening
stellar contaminant are not incorporated -- these are too uncertain to
model with any accuracy.  The mass function of the lenses is that
deduced by Gould, Bahcall \& Flynn (1997) from analysis of star count
data in the Groth Strip. The velocity distribution is a trivariate
Gaussian with semi-axes $2:1:1$ about the circular speed. The
underlying histogram is the four-year dataset of the MACHO 
group (Alcock et al. 1997a, Axlerod 1997).}

\end{figure}

\begin{figure}

\rotate[r]{
               \epsfxsize 0.7\hsize 
               \epsffile{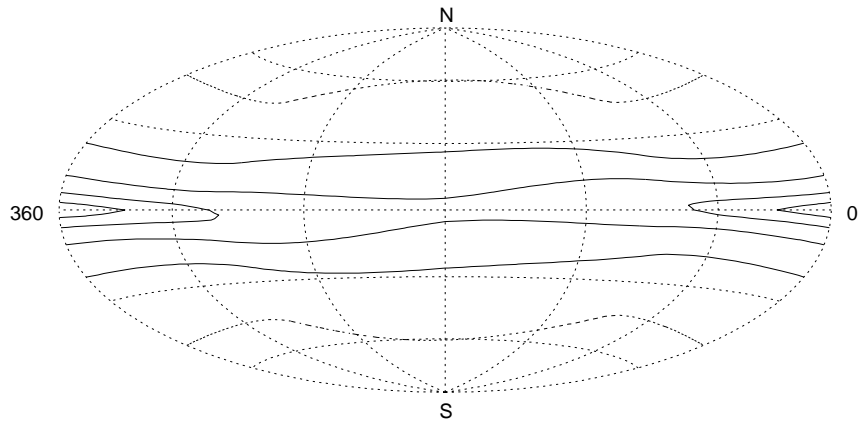}
}
\caption{The contours of surface brightness of our warped and flaring
disk model viewed in a Hammer-Aitkof projection towards the
anti-center.  This figure is obtained by projecting the disk density
given in equation (1) assuming a constant mass-to-light ratio of 3.
At the location of the LMC ($\ell = 280^\circ, b = -33^\circ$), the
surface brightness is smooth, and therefore would have been removed by
de Vaucouleurs prior to constructing his LMC isophotal map. }

\end{figure}

\end{document}